# Heavy fermions, mass renormalization and local moments in magic-angle twisted bilayer graphene via planar tunneling spectroscopy

Zhenyuan Zhang[1,*], Shuang Wu[1,*], Dumitru Călugăru[2,3], Haoyu Hu[2,4], Takashi Taniguchi[5], Kenji Wanatabe[6], Andrei B. Bernevig[2,4,7] and Eva Y. Andrei[1,†]

[1]Department of Physics and Astronomy, Rutgers University, Piscataway, New Jersey, 08854, USA

[2]Department of Physics, Princeton University, Princeton, New Jersey, 08544, USA

[3]Rudolf Peierls Centre for Theoretical Physics, University of Oxford, Oxford OX1 3PU, United Kingdom.

[4]Donostia International Physics Center (DIPC), Paseo Manuel de Lardizabal. 20018, San Sebastián, Spain

[5]International Center for Materials Nanoarchitectonics, National Institute for Materials Science, 1- 1 Namiki, Tsukuba 305-0044, Japan

[6]Research Center for Functional Materials, National Institute for Materials Science, 1-1 Namiki, Tsukuba 305-0044, Japan

[7]IKERBASQUE, Basque Foundation for Science, 48013 Bilbao, Spain

* These authors contributed equally to the work.

† Corresponding author. Email: andrei@physics.rutgers.edu

**Topological heavy fermion models[1-5] describe the flat bands in magic-angle twisted bilayer graphene (MATBG) as arising from the hybridization between localized flat-band orbitals (*f*-electrons) and nearly-free conduction bands (*c*-electrons). The interplay between these *f*-electrons and *c*-electrons is theorized to give rise to emergent phenomena, including unconventional superconductivity[6-8], non-Fermi liquid behavior[9-11], and topologically nontrivial phases[12-14]. However, the fundamental properties of *f*- and *c*-electrons, such as their respective heavy and light effective mass and their properties under strain, need experimental verification. Here we report on the electronic inverse**

**compressibility, effective mass, and entropy of MATBG, obtained from planar tunneling spectroscopy. Our results include the observation of electron mass renormalization, found to be consistent with the topological heavy fermion model prediction of heavy charge-one excitations away from integer fillings. Importantly, we present entropic evidence for 4-fold and 8-fold degenerate isospin local moment states emerging at temperatures of 10K and 20K, respectively, consistent with the entropy of 8 heavy-fermions flavors energetically split by the sample strain.**

Twisted bilayer graphene near the magic-angle of 1.1° (MATBG) features two 4-fold degenerate flat bands near charge neutrality, which are isolated from the high-energy remote bands. Owing to the quenched kinetic energy and degeneracy of these flat bands, electron-electron interactions and isospin correlations (spin, valley, sublattice) lead to significant band reconstruction[15, 16] observed as a cascade of electronic transitions at integer moiré fillings [11, 13, 14, 17-24] Recent Kondo lattice models [5, 25-29] not only align with the low-temperature heavy fermion behavior observed in MATBG [2, 30], but also predict the formation of isospin local moments and a cascade of spectral weight reorganizations at higher temperatures that have thus far eluded experimental confirmation.

Planar tunneling spectroscopy (PTS) [31-36] provides insights into key characteristics of the electronic properties including energy gaps, mid-gap states, and quasiparticle excitations. PTS holds several advantages over traditional scanning tunneling spectroscopy (STS), including an ultra-stable tunneling junction, high sensitivity, high energy resolution and ready access to ultra-low temperatures and magnetic fields in standard dilution refrigerator setups. Here, we employed PTS to measure the density of states (DOS), the electronic inverse compressibility (d$\mu$/d$n$) and entropy in MATBG, leading to the observation of heavy electron mass renormalization and entropic evidence of isospin states. Our measurements reveal distinct ground states at various fillings and uncover mass-renormalization transitions from heavy to light fermions across integer fillings, that persist up to high temperatures (~70 K). These transitions can be effectively described by a Kondo lattice model featuring localized *f*-electrons at AA sites coexisting with itinerant *c*-electrons[28, 29]. By examining the temperature and magnetic field dependence of the DOS and $d\mu/dn$, we uncovered local moment states emerging at finite fields and temperatures. The data reveal two entropy plateaus providing entropic evidence for 4-fold and 8-fold degenerate isospin local moment states emerging at 10K and 20K, respectively, consistent with the entropy of 8 heavy fermions flavors that are energetically split by the sample strain.

**Planar tunneling spectroscopy**

The planar tunneling device (Fig 1a) comprises a thin (2 nm) hexagonal boron nitride (h-BN) flake serving as the tunneling barrier, a graphite ribbon serving as the tunneling probe, and a top gate separated from the MATBG sample by a thicker (30 nm) hBN layer (Methods). The small, constant and stable probe-sample distance produces an ultra-stable tunneling junction leading to high energy resolution and provides access to the inverse compressibility. The twist angle, $1.06° \pm 0.02°$, is determined by calibrating the carrier density *n* using high magnetic field Landau levels (LLs) and Chern insulators (CIs) (Methods). Fig.1b shows the bias and gate voltage dependence ($V_b$ and $V_g$, respectively) of the differential conductance $dI/dV_b$ measured at 0.3 K. Near $V_b = 0$, we observe a gap-like feature with vanishingly low $dI/dV_b$ across all values of $V_g$. This feature, commonly observed in graphene-based tunneling, reflects the momentum mismatch between the carriers emanating from a metallic tip (zero momentum) and the low energy states in graphene with momentum $|\vec{K}| = \frac{4\pi}{3a} = 1.7 \text{Å}^{-1}$, where *a* = 0.246 nm is graphene's lattice spacing. Tunneling into graphene can proceed via an inelastic process involving a ~60meV flexural phonon that bridges the momentum mismatch and gives rise to the gap feature known as the "phonon gap" [33, 37-39], or alternatively by using an atomically sharp STM tip which naturally emits electrons with large momenta[40, 41]. Here we use graphite as the tunneling probe, with its lattice orientation misaligned with that of the TBG sample by ~18° as measured by AFM imaging (SI). This leaves a momentum mismatch of $|\vec{K_g} - \vec{K_{TBG}}| = \Delta k \sim 0.53$ Å$^{-1}$ that requires a 20 meV acoustic phonon (ZA branch) to open a channel of inelastic tunneling (Fig. 1d), resulting in the phonon gap observed as a dark central strip in the d$I$/d$V_b$ map (Fig. 1b). From the d$I$/d$V_b$ spectrum and its derivative, $d^2I/dV_b^2$, at $V_g$ = -10V (Fig. 1c) we note that the phonon gap is symmetric about $V_b = 0$ and is flanked by sharp peaks in $d^2I/dV_b^2$ representing the phonon-mediated inelastic tunneling channels at ±22 mV[37], consistent with the required bridging phonon of momentum 0.53 Å$^{-1}$ (Fig. 1d). In the presence of a phonon gap the Fermi level, $E_F$, is pushed from zero energy to the energy at which inelastic tunneling sets in. We thus have $E_F = \mu(n) \sim \pm 22$ mV, where $\mu(n)$, the chemical potential measured with respect to the charge neutrality point (CNP), is identified with $E_F$.

The red and green dashed lines in Fig. 1b marking the peaks in the differential conductance, track the gate dependence of the centers of the conduction and valence bands respectively. When the active MATBG bands are either empty ($V_g$ < -8V) or full ($V_g$ > +8V), the two bands, which are 16 mV apart and parallel to each other, are strongly doping dependent. For partial filling (-8V < $V_g$ < +8V) and away from the CNP ($V_g$ ~0), they are independent of doping (horizontal dashed lines) and feature a ~15 mV splitting upon crossing the Fermi level (Fig. 1e). Near the CNP the bands develop a slope indicating the presence of a gap consistent with STS and SET measurements [14, 17-19, 21-24, 42].

Next, we discuss the dark tilted lines marked by orange arrows in Fig.1b, where $dI/dV_b$ is suppressed up to high bias voltages. Their slope which reflects the carrier density dependence on both $V_g$ and $V_b$, is given by the ratio $C_g/C_b \sim 1/27$ where $C_g$ and $C_b$ are the sample to gate-electrode and sample to bias-electrode capacitances, respectively. The total induced charge is thus $ne = C_g V_g - C_b V_b$, where $e$ is the electron charge. This differs from the approximation used in transport and STM/STS measurements, where $ne \approx C_g V_g$ because $C_b$ is negligibly small (due to small probe area in STS and large probe sample distance in transport). The sheared $dI/dV_b$ in Fig.1e presents the same $dI/dV_b$ data as in Fig. 1b, but plotted as a function of $V_b$ and the carrier density per moiré cell, $n/n_0$, where $n_0$ corresponds to one carrier per moiré cell. This eliminates the explicit $V_b$ dependence of $n$, so that the dark lines become vertical appearing close to moiré fillings $n/n_0 = \pm 4, 0$ [31, 35, 43]. In Fig. 1f showing individual spectra at integer moiré fillings we marked the conduction and valence band peaks by red and green arrows respectively.

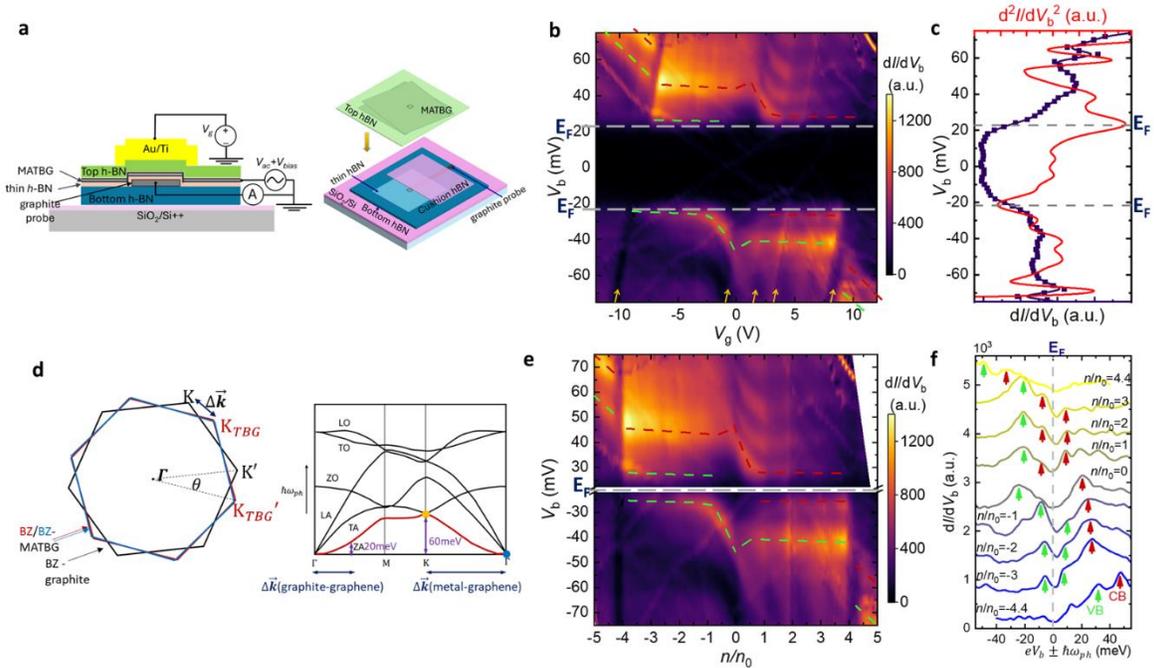

**Fig.1 Device structure and $dI/dV_b$ map. a**, Schematic diagram of the device structure (Left) and stacking order of the van der Waals heterostructures (Right), showing the planar tunneling measurement configuration with a graphite probe. The AC voltage can be applied to either the bias voltage (as shown in the diagram) or the gate voltage. See Methods for details. **b**, $dI/dV_b$ of 1.08° MATBG ($V_{ac} = 0.8$ mV, 11.07 Hz) as a function of $V_b$ and $V_g$. Red and green dashed lines mark the conduction and valence flat bands. Dark tilted lines indicating the suppressed dI/dV$_b$ are marked by orange arrows. **c**, $dI/dV_b$ and $d^2I/dV_b^2$ at $V_g = -10$ V. Grey dashed lines mark the onset of inelastic tunneling channels, and the Fermi level in the spectrum is around $V_b = \pm22$ mV. This onset voltage is independent of $V_g$. **d**, Left: Momentum-space geometry of MATBG and graphite probe. There is a momentum mismatch between the sample and the probe due to an 18° orientation difference which can be bridged at a higher energy level by absorbing or emitting a phonon. Right: The scheme of the phonon spectrum of graphite. The blue solid dot represents electrons in the metal probe. The orange solid dot represents electrons in TBG. The momentum and energy required from the phonon are labeled for tunneling between the metal probe and TBG, and for tunneling between the graphite probe and TBG on the right and left, respectively. **e**, Same as **b**, but sheared to show filling dependence. The phonon-induced gap feature near $V_b = 0$ mV is omitted. Red and green dashed lines mark the conduction and valence flat bands. Suppressed $dI/dV_b$ is observed at $n/n_0 = \pm4$. **f**, $dI/dV_b$ spectra at varied fillings. The peaks indicating conduction band (CB), and valance band (VB) are marked by red and green arrows.

**Electronic inverse compressibility and chemical potential of correlated gaps and Chern insulators**

To analyze the data, we model the tunneling current using the usual assumptions of a constant probe DOS and energy-independent (but filling-dependent) transmission coefficient *T(n)* [31, 35, 43]:

$$I \propto T(n) \int_{\mu(n)}^{\mu(n)+eV_b} \rho(\epsilon)d\epsilon, \quad (1)$$

where $\rho(\epsilon)$ is the sample DOS. In order to access $d\mu/dn$, we divide *I* by the transmission coefficient background, $T(n) \propto \exp\left(0.258 \cdot \frac{n}{n_0}\right)$, arising from the dependence of the barrier on carrier density (SI and Extended Figure 1a). Taking the derivative of the normalized tunneling current $I_n=I/T(n)$ with respect to $V_b$, and recalling that $E_F = \mu(n)$, we obtain the differential conductance (Methods):

$$\left.\frac{dI_n}{dV_b}\right|_{V_b} \propto \rho(E_F + eV_b) - \frac{d\mu(n)}{dn}\frac{C_b}{e^2}\left(\rho(E_F + eV_b) - \rho(E_F)\right) \qquad (2)$$

The first term, $\rho(E_F + eV_b)$, represents the sample DOS similar to that measured in STS, while the second term, which contains $\frac{d\mu}{dn}$, is usually neglected in standard STS due to the small tip-sample capacitance $C_b$. In contrast to STS, the large value of $C_b = \frac{e\partial n}{\partial V_b} \sim 1.3\ \mu\text{F/cm}^2$ in our planar tunneling device makes it possible to gain access to d$\mu$/d$n$ from the measured tunneling differential conductance.

The ultra-stable tunneling junction of the planar tunneling structure makes it possible to measure d$I$/d$V_g$, in addition to the standard d$I$/d$V_b$. Based on these two measurements we developed a method to extract d$\mu$/d$n$ from the tunneling data (Methods):

$$\left.\frac{d\mu(n)}{dn}\right|_{V_b} \approx \frac{e^2 \frac{dI_n}{dV_g}}{C_g \frac{dI_n}{dV_b} + C_b \frac{dI_n}{dV_g}}. \qquad (3)$$

A similar expression was used in SET measurements[22], but the probing bias voltage contribution to doping was not included presumably due to a much larger probe-sample distance (16nm). An example of extracting d$\mu$/d$n$ on the electron side at $V_b$ = -42mV and 4K curves is shown in Extended Fig.1b, c. We note that the d$\mu$/d$n$ peaks corresponding to the expected band gaps at fillings of ±4 are shifted away by about 0.2 (Extended Fig.1c). This discrepancy is due to approximating the capacitance by its geometric contribution, $C_{geo} \approx C_g$ when calculating the carrier density. In fact, the geometric capacitance and quantum capacitance, $C_q$, are connected in series, so that the actual capacitance relevant for the doping calculation is $C = \left(\frac{1}{C_{geo}} + \frac{1}{C_q}\right)^{-1} \approx \left(\frac{1}{C_g} + \frac{1}{C_q}\right)^{-1}$. While $C_g$ depends only on sample-gate distance and is a known constant, $C_q$ is proportional to the DOS. Therefore, when the sample enters insulating states at band edges, the effect of the quantum capacitance becomes stronger, and the filling calculation must be adjusted. After applying this correction (SI), the d$\mu$/d$n$ peaks marking the band edges separating the flat bands from remote bands, appear at $n/n_0 = \pm 4$, as shown in Fig.2a where we plot the filling dependent d$\mu$/d$n$ curves at 4K and at 0.3K. Away from the band edges the filling correction is negligibly small (SI). We note that within the flat bands d$\mu$/d$n$ reveals strong filling-dependent variations, changing from low or even negative values on the lower-doping side of integer fillings to high peaks on the higher-doping side. This dip-peak sequence is more pronounced on the electron side than on the hole side, consistent with earlier reports[17]. Similar non-monotonicity in the filling dependence of the flat bands observed with other techniques, was attributed to a sequence of Fermi surface reconstructions [11, 13, 14, 17-23]. As $n/n_0$ approaches each non-zero integer filling from the lower doping side, d$\mu$/d$n$ turns negative, indicating the emergence of a more stable lower energy state. Upon crossing an integer filling, d$\mu$/d$n$ increases until it peaks just below the next integer filling. The d$\mu$/d$n$ peak following a negative dip resembles the correlation gap seen in Mott insulators [44], though here it is likely related to the heavy to light electron transition as discussed below. An exception occurs at 0.3K, where a sharp d$\mu$/d$n$ peak precisely at $n/n_0 = 2$ accompanied by a dip below the same filling (inset Fig.2a) [20, 42], reveals the emergence of a correlation induced gap and possibly symmetry breaking. This gap feature is consistent with the strong resetting of the Hall number at this filling observed [13] when six of the eight bands (two bands for each spin and valley flavor) are filled while the other two remain empty. The absence of a similar peak at odd integer fillings (1 and 3) is consistent either with a gapless ordered state (such as the incommensurate Kekulé spiral state (IKS)[45] at filling +1, or with a symmetric state [5, 46].

The chemical potential, $\mu$, relative to the CNP, is obtained by integrating d$\mu$/d$n$ over the carrier density, $\mu(n/n_0) = \int_0^{n/n_0} \left(\frac{d\mu}{dn}\right) dn$. From the filling dependence of $\mu(n/n_0)$ shown in Fig.2b we observe a total bandwidth of 26meV as well as a 10meV gap at the CNP which is determined from the jump in the chemical potential (red arrow). At filling $n/n_0 = 2$, the jump in the chemical potential is due to a combination of the correlation induced gap and band reconstruction. An estimate of this gap obtained from the change in chemical potential across the filling range (1.9 to 2.1) (Fig. 2b inset), $\Delta = 1.3 \pm 0.14$ meV, is consistent with the gap value measured in transport [13]. Fig.2c shows the magnetic field dependence (0 T to 6 T) of $d\mu/dn$ in the electron sector at 0.3 K. (hole sector discussed in the SI). The pattern reveals a Landau fan similar to that seen in transport measurements, but here the gaps between LLs or Chern bands correspond to $d\mu/dn$ maxima rather than to the $R_{xx}$ minima in transport. Near the CNP, we observe LLs (Extended Figure 2) whose 4-fold degeneracy is lifted above 3.5 T, revealing a full LL fan emanating from the $0^{th}$ LL and half fans emerging from fillings 1, 2 and 3. The latter replicate those identified in transport measurements as Chern insulators states [13]. Their Chern numbers are extracted using the Streda formula: $C = \nu = (n/n_0 - s)/(\phi/\phi_0)$, where $s = \pm 1, \pm 2, \pm 3$ is the band index and $s \cdot \nu > 0$ [13, 14, 21, 47-49] The peak associated with the correlation-induced gap remains fixed at $n/n_0 = 2$, for all magnetic fields below 5.5 T (yellow dashed line). In Fig.2d, the filling dependent chemical potential curves illustrates the evolution of LLs and CIs with magnetic field.

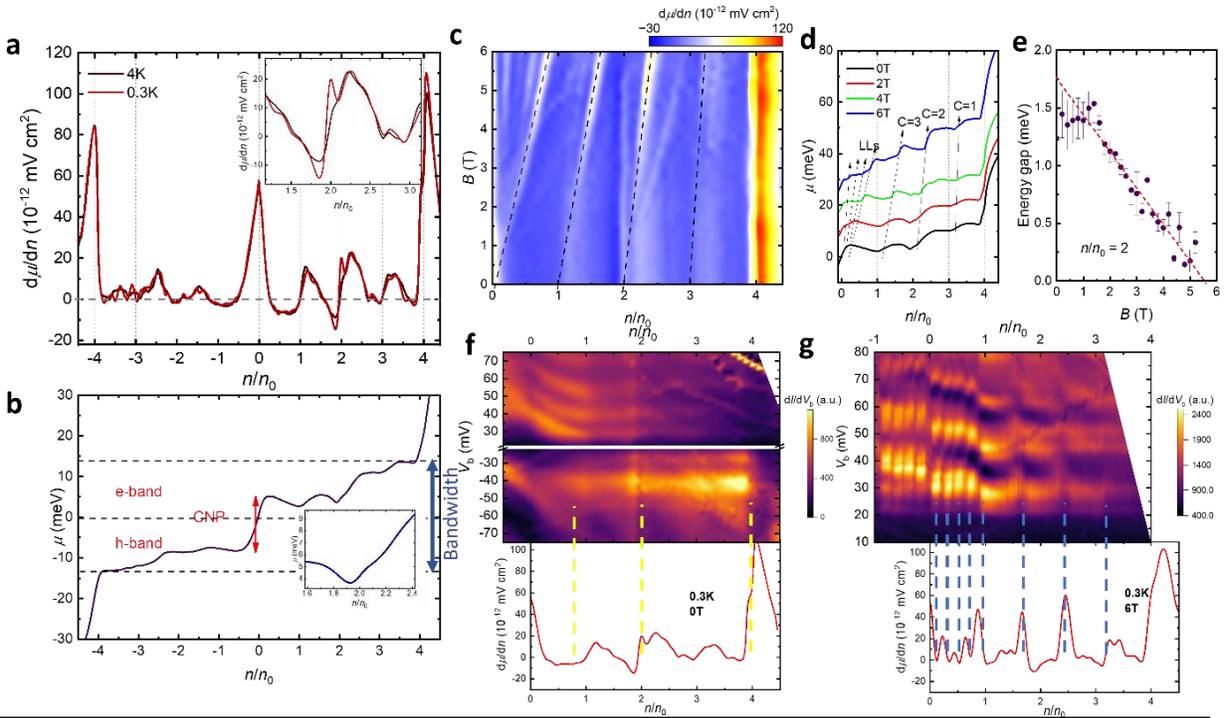

**Fig. 2 Electronic inverse compressibility and magnetic field dependence. a**, Extracted d$\mu$/d$n$ at 0.3 K and 4 K from d$I$/d$V_b$ and d$I$/d$V_g$ at $V_b = \pm 42$ mV. Inset, zoomed-in curves near filling of 2 showing the opening of a gap right at filling of 2 below 4K. **b**, Chemical potential $\mu$ integrated from d$\mu$/d$n$ in **a** as function of filling at 0.3K. Inset, zoom-in around the filling of 2. **c**, d$\mu$/d$n$ map on the electron side as a function of filling and perpendicular magnetic field. The Landau filling factor and Chern number can be extracted from the response of d$\mu$/d$n$ peaks of LL and Chern gaps labeled by dashed lines. **d**, Chemical potential (stacked) on the electron side at various perpendicular magnetic fields. Arrows mark the LL and Chern gaps developing under magnetic fields. All the degeneracies of the 0$^{th}$ LL are lifted under high magnetic fields. **e**, Gap (jump in chemical potential) at filling of 2 versus perpendicular magnetic field. The red dashed line is a linear fit for $B > 1.6$ T yielding a $g$-factor of 5.4. **f**, Top panel: d$I$/d$V_b$ map at 0.3 K and zero magnetic field. Bottom: d$\mu$/d$n$ at 0.3 K and zero magnetic field. The d$\mu$/d$n$ peaks at $n/n_0 = 0, 2, 4$ align well with dark vertical lines in the top panel indicating low DOS. **g**, Same as **f**, but under a perpendicular magnetic field of 6 T. Peaks in d$\mu$/d$n$ caused by LLs and CIs align with dark vertical lines in the d$I$/d$V_b$ map.

The magnetic field dependence of the gap size is shown in Fig.2e. Above 1.5T, the gap decreases nearly linearly with field, suggesting a spin-unpolarized ground state, consistent with the intervalley coherent spin singlet observed in transport measurements [13]. A linear fit of $\Delta = \Delta_0 - g\mu_B B$ yields a $g$-factor of $5.4 \pm 0.2$, where $\Delta_0 = 1.75$ meV, and $\mu_B$ is the Bohr magneton. Below 1.5 T, we note that the gap size remains roughly constant. Fig. 2f and Fig. 2g showing a side-by-side comparison of the $dI/dV_b(V_b, n/n_0)$ and $d\mu/dn$ $(n/n_0)$ without and with magnetic field respectively, illustrates the direct correspondence between the vertical dark lines in the $dI/dV_b$ map and the $d\mu/dn$ peaks.

**Heavy fermion and mass renormalization**

Next, we use the $\frac{d\mu}{dn}$ data to obtain the effective mass and its filling dependence. The cyclotron effective mass is defined as $m^* = \frac{\hbar^2}{2\pi}\frac{\partial A}{\partial E}$, where $\hbar$ is the reduced Planck's constant, and $A$ is the Fermi surface area. In 2D systems in the absence of correlation effects the effective mass is given by

$$m^* = \frac{2\pi\hbar^2}{g'}\left(\frac{d\mu}{dn}\right)^{-1}, \tag{4}$$

where $g'$ is the electron degeneracy. While this expression is valid for weakly-correlated excitations with doping-independent degeneracy, we use it to qualitatively estimate the carriers' effective mass, assuming a doping-independent band degeneracy $g' = 4$ (corresponding to the two spin and two valley degrees of freedom). Fig.3a shows d$\mu$/d$n$ and m*/$m_e$ ($m_e$ bare electron mass) extracted using equation 4 at 4 K over the full range of band fillings. Remarkably, close to positive integer fillings, n/$n_0$ =1, 2, 3, m*/$m_e$ undergoes dramatic swings in mass renormalization, ranging from large values (0.6 $m_e$ or higher) just below integer filings, to values as low as ~0.06 $m_e$ just above integer fillings which are comparable to values in bilayer graphene (0.03-0.05 $m_e$)[50]. This mass asymmetry of the carriers is consistent with recent Seeback experiments observing heavy (light) excitations below (above) positive integer fillings[51]. The temperature dependence of m*/$m_e$ shown Fig. 3b, reveals that the strong mass renormalization in the flat band around integer fillings persists up to ~ 50K. We note that around the CNP (where both the hole and electron excitation are light) as well as in the remote bands, m*/$m_e$ can be calculated directly from the LL fans (Methods).

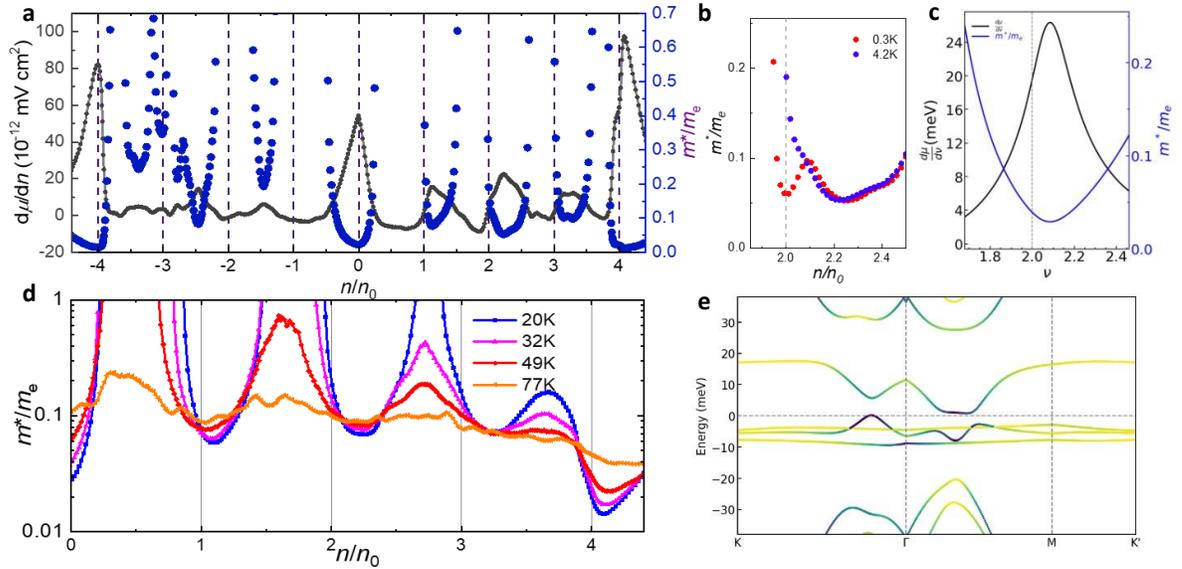

**Fig. 3 Heavy fermion and mass renormalizations at integer fillings. a**, $d\mu/dn$ and cyclotron effective mass m* in units of electron mass $m_e$ as a function of $n/n_0$ at 4 K. Upon approaching each positive integer filling 1, 2, and 3, m* drops from high to low, reaching a minimum at filling of $n/n_0 + \delta$. **b**, Cyclotron effective mass m* in units of electron mass me around filling of 2 at 0.3K and 4K **c**, Theoretical calculation of the $d\mu/dv$ and m*/$m_e$ around filling of 2 at T=20K and compressive heterostrain ε=0.001 (see SI) here $v \equiv n/n_0$ **d**, Cyclotron effective mass m* in units of $m_e$ above 20 K. Prominent cascades of transitions from heavy to light across integer fillings survive until ~77 K. A factor of 10 is determined in the m* renormalization. **e**, Theoretically calculated band structure with topological heavy-fermion model (ε=0.001, see SI): *f*-orbitals (yellow) and *c*-orbitals (blue). With hybridization, charge-one excitation bands are formed having both f- and c- character.

The filling dependence of m*/$m_e$ observed here is consistent with both symmetry-broken[1, 2, 52, 53] and symmetric[29, 46, 54] phases of MATBG. The effective mass asymmetry naturally follows from the topological Heavy Fermion model[2, 25, 27-29, 46, 55, 56] and related models[3-5, 54] for MATBG. In these models, the active MATBG bands emerge from the hybridization between localized, dispersionless and strongly correlated *f*-electrons – experimentally observed via STM/STS to occupy the AA moiré lattice[18, 57] – and itinerant, weakly correlated *c*-electrons. At positive, nonzero integer fillings, in both symmetric and symmetry-broken states, hole excitations predominantly exhibit *f*-character, whereas electron excitations possess *c*-character. This distinction naturally leads to significant differences in d$\mu$/d$n$ and $m^*$ around integer fillings, as doping selectively excites carriers of either *f*- or *c*-character. Specifically, electron excitations remain light, while hole excitations are significantly heavier around positive, nonzero integer fillings.

To illustrate the asymmetry of charge-one excitations, we consider the incommensurate Kekulé spiral (IKS) state[45], a symmetry-broken phase proposed in the presence of strain and identified through STM experiments[58]. The wave function of this state was analytically determined in Ref.[59] At $n/n_0 = +2$, this state is theoretically predicted to be gapped. Our measurements reveal a sharp peak in d$\mu$/d$n$ precisely at the integer filling, supporting the presence of a gap emerging at the lowest temperatures. Moreover, the effective masses of electron and hole excitations show good agreement with our experimental findings at the lowest temperatures around this filling (Fig.3b,3c). The distinct character of electron and hole excitations near integer fillings is further evident in Fig.3d. At other positive integer fillings at low temperatures, as well as at all positive integer fillings at higher temperatures, the effective mass asymmetry persists, showing its robustness across both symmetric and symmetry-broken phases of the system. The displaced peak in d$\mu$/d$n$ from integer fillings is consistent with either a gapless symmetry-broken state, such as the IKS state at $n/n_0 = +1$,[45, 59] or a fully symmetric state[54].

**Entropic evidence of local moment states**

Figure 4a shows the temperature dependence of d$\mu$/d$n$ on the electron side, where the cascade of electronic transitions persists up to ~70 K. Using Maxwell's relation $\frac{\partial S}{\partial n} = -\frac{\partial \mu}{\partial T}$, and the assumption that the entropy (S) is zero at full filling[11, 42], we calculate the filling dependence of the entropy, S(n), from the measured chemical potential (Methods). The resulting S(n) curves on the electron side are shown in Fig. 4b. Figure 4c displays the entropy at fillings 0, 1, 2, and 3 over a temperature range of 4.2 K to 36 K. As reported in previous studies, the entropy increases with rising temperature, as it should[11, 42]. However, the key novel observation from our high-resolution entropy data is the appearance of two plateaus: one near 10 K and another near 20 K, which aligns well with the formation of solid-like local moment states that are temperature-independent[25]. According to Kondo lattice models[5, 25-29], at relatively high temperatures, the $c$- and $f$- fermions can be treated as decoupled. In the resulting so-called zero-hybridization limit (SI)[60], electrons occupy the $f$-fermion states at integer fillings, with the $c$-electron sector being half-filled. This results in a localized solid-like $f$-electron lattice at the AA moiré sites, each hosting random isospin local moments (comprising spin, valley, and sublattice degrees of freedom). The presence of two plateaus, rather than one, suggests that the degeneracy of these local moments changes between 10K and 20K. To test this hypothesis, our filling-dependent entropy measurements between 5K to 24K are compared with a simple model that considers integer $f$-electrons with random local moments (Fig.4b). Around 20K, corresponding to the high-temperature plateau, the entropy matches $k_B \ln \binom{8}{n/n_0 + 4}$, accounting for the microstates with a degeneracy of 8. Around 10K, at the low-temperature plateau, the entropy aligns with $k_B \ln \binom{4}{|n/n_0|}$, corresponding to a degeneracy of 4. This model of changing degeneracy is further supported by the fact that the entropy at the CNP saturates to zero below 12 K, as also observed in SET measurements of the inverse compressibility.

The change in degeneracy could result from, and is consistent with, that from broken symmetries, such as those caused by strain[61], which is known to exist in these systems. Fig.4d shows the total entropy of $f$- and $c$-electrons at integer fillings, calculated using the zero-hybridization limit of the topological heavy-fermion model with strain (SI)[2, 61]. This calculation agrees well with the two observed entropy plateaus. Since the simplified model ignores f-c hybridization and correlations between AA sites, the calculated entropy does not approach zero at low temperatures, as expected for heavy Fermi liquid or for correlated

symmetry-broken insulators[25, 61]. We note that based on the calculation, the entropy increasing

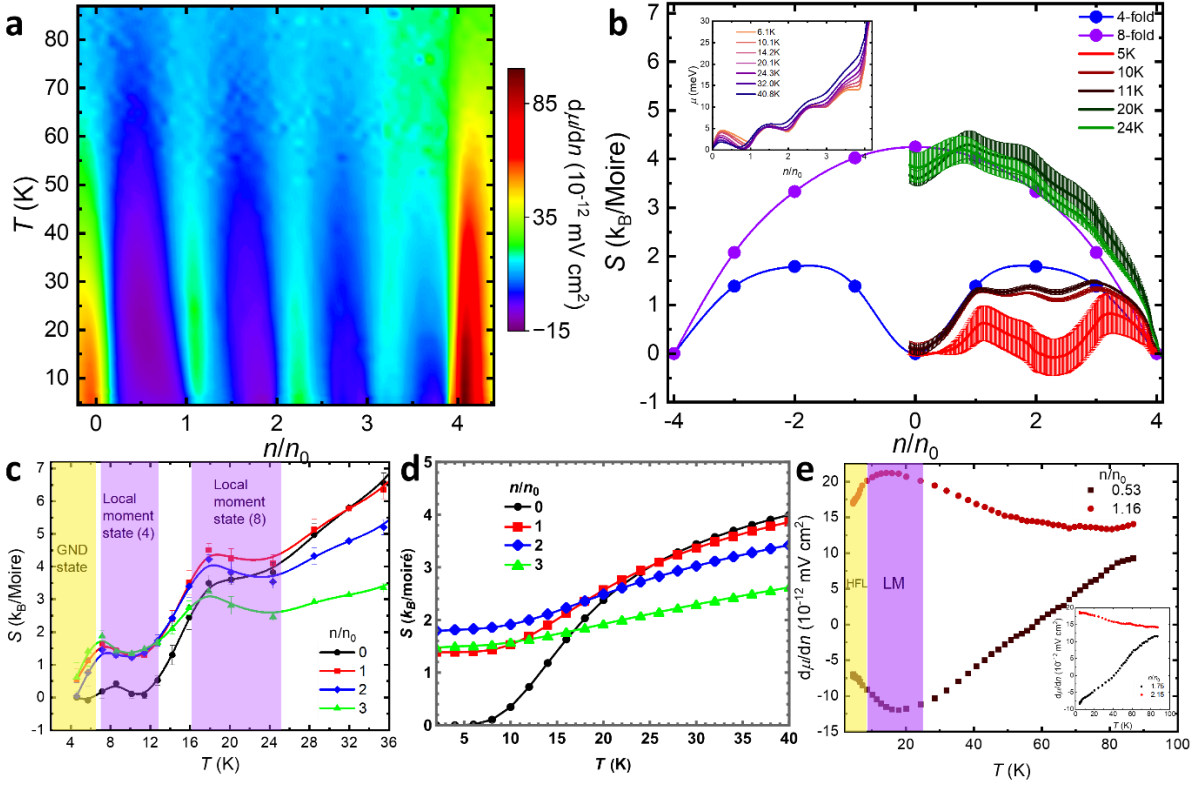

**Fig. 4 Temperature dependence of inverse compressibility and entropy. a**, d$\mu$/d$n$ map on the electron side as a function of filling and temperature. The cascade transitions persist up to 70 K. **b**, The electronic entropy $S$ as a function of filling at various temperatures. Blue and purple dots connected with dashed lines are calculated by counting all possible microstates with degeneracy of 4 and 8, respectively. Inset, chemical potential $\mu$ calculated by integrating d$\mu$/d$n$ at various temperatures. **c**, The electronic entropy $S$ at non-negative integer fillings. The entropy is obtained by solving Maxwell's relation for every adjacent three temperature data points. Two plateaus are observed near 10 K and 20 K. **d**, Total entropy at integer fillings calculated in the atomic limit of the topological heavy-fermion model with strain. The Hubbard interaction energy $U_1$ of 23meV and the compressive heterostrain $\varepsilon$=0.001 are set to match the experimental data (see SI). Below and above 15 K, the entropy shows two plateaus corresponding to the degeneracy of 4 and 8, in good agreement with the measured entropy. Note that in the atomic limit, the $f$-electrons at different lattice sites are decoupled from one another and also decoupled from the $c$-electrons, which is not valid for the real ground states at low temperatures below 6 K. **e**, Non-monotonic temperature dependence of d$\mu$/d$n$ near $n/n_0 = 1$. Inset, the temperature dependence of d$\mu$/d$n$ near $n/n_0 = 2$ as a comparison. The non-monotonic behavior can also be seen from the color map **a**, which is linked to the Pomeranchuk-like effect at $n/n_0 = 1$.

with temperature may be related to the phonon or low-energy charge excitations (SI) [62].

The entropy plateaus near 10K and 20K (Fig.4c) provide direct evidence of isospin local moment states in MATBG at relatively high temperatures (8K to 24K). Another significant feature is the much lower entropy at a filling of 2 compared to fillings of 1 and 3 at 5 K (Fig.4b), which is consistent with the gapped ground state at filling 2, as seen in the d$\mu$/d$n$ peak below 4 K (inset, Fig.2a).

**Conclusions**

The results demonstrate the evolution of flat bands and the emergence of Landau levels and Chern insulators in MATBG, as observed through planar tunneling spectroscopy. Electronic inverse compressibility, d$\mu$/d$n$, extracted from differential tunneling conductance (d$I$/d$V_b$ and d$I$/d$V_g$), reveals a pronounced and robust cascade of transitions between heavy and light fermions across integer fillings. Two distinct entropy plateaus, around 10K and 20K, suggest the presence of two local moment phases with degeneracy of 4 and 8, respectively. Both the d$\mu$/d$n$ and the entropy results are well explained by Kondo lattice models[5, 25-29], which account for the interplay between heavy localized $f$-electrons and light itinerant $c$-electrons. The proposed scenario of melting local moment states will be further investigated through additional theoretical and experimental studies.

**Data Availability**

The data that support the findings of this work are available from the corresponding author upon reasonable request.

**Author Contributions**

Z.Z., S.W., and E.Y.A. conceived and designed the experiment, carried out low-temperature tunneling measurements, and analyzed the data. S.W. and Z.Z. fabricated the planar tunneling twisted bilayer graphene devices. D.C., H.H., and B.A.B performed theoretical simulations, T.T.

and K.W. synthesized the h-BN crystals. Z.Z., S.W., D.C., H.H., B.A.B and E.Y.A. wrote the manuscript.

**Acknowledgments**

We thank Gautam Rai, Jonah Herzog-Arbeitman, Andrew Millis, Chung-Hou Chung, Piers Coleman, Liam LH Lau, Zhi-Da Song, for useful insights and discussions. Support from DOE-FG02-99ER45742 (EYA, ZZ) and from Gordon and Betty Moore Foundation GBMF9453 (SW, ZZ, EYA) is gratefully acknowledged. D.C. thanks the hospitality of the Donostia International Physics Center, at which this work was carried out. D.C. also gratefully acknowledges the support provided by the Leverhulme Trust. H.H. was supported by the European Research Council (ERC) under the European Union's Horizon 2020 research and innovation program (Grant Agreement No. 101020833), as well as by the IKUR Strategy under the collaboration agreement between Ikerbasque Foundation and DIPC on behalf of the Department of Education of the Basque Government. B.A.B. acknowledges support from the DOE Grant No. DE-SC0016239. K.W. and T.T. acknowledge support from the Element Strategy Initiative conducted by the MEXT, Japan, grant JPMXP0112101001 and JSPS KAKENHI grants 19H05790, JP20H00354, and 21H05233.

**Reference**

1. Kang, J., B.A. Bernevig, and O. Vafek, *Cascades between Light and Heavy Fermions in the Normal State of Magic-Angle Twisted Bilayer Graphene.* Physical Review Letters, 2021. **127**(26): p. 266402.
2. Song, Z.D. and B.A. Bernevig, *Magic-Angle Twisted Bilayer Graphene as a Topological Heavy Fermion Problem.* Physical Review Letters, 2022. **129**(4).
3. Shi, H. and X. Dai, *Heavy-fermion representation for twisted bilayer graphene systems.* Physical Review B, 2022. **106**(24): p. 245129.
4. M. Haule, E.Y.A., K. Haule, *The Mott-semiconducting state in the magic angle bilayer graphene.* arXiv:1901.09853, 2019.

5.  Datta, A., et al., *Heavy quasiparticles and cascades without symmetry breaking in twisted bilayer graphene.* Nature Communications, 2023. **14**(1).
6.  Cao, Y., et al., *Unconventional superconductivity in magic-angle graphene superlattices.* Nature, 2018. **556**: p. 43.
7.  Yuan Cao, D.R.-L., Jeong Min Park, Fanqi Noah Yuan, Kenji Watanabe, Takashi Taniguchi, Rafael M. Fernandes, Liang Fu, Pablo Jarillo-Herrero, *Nematicity and Competing Orders in Superconducting Magic-Angle Graphene.* arXiv:2004.04148, 2020.
8.  Lu, X., et al., *Superconductors, orbital magnets and correlated states in magic-angle bilayer graphene.* Nature, 2019. **574**(7780): p. 653-657.
9.  Cao, Y., et al., *Correlated insulator behaviour at half-filling in magic-angle graphene superlattices.* Nature, 2018. **556**: p. 80.
10. Cao, Y., et al., *Strange Metal in Magic-Angle Graphene with near Planckian Dissipation.* Physical Review Letters, 2020. **124**(7): p. 076801.
11. Rozen, A., et al., *Entropic evidence for a Pomeranchuk effect in magic-angle graphene.* Nature, 2021. **592**(7853): p. 214-+.
12. Jeong Min Park, Y.C., Kenji Watanabe, Takashi Taniguchi, Pablo Jarillo-Herrero, *Flavour Hund's Coupling, Correlated Chern Gaps, and Diffusivity in Moiré Flat Bands.* arXiv:2008.12296, 2020.
13. Wu, S., et al., *Chern insulators, van Hove singularities and topological flat bands in magic-angle twisted bilayer graphene.* Nature Materials, 2021. **20**(4): p. 488-494.
14. Nuckolls, K.P., et al., *Strongly correlated Chern insulators in magic-angle twisted bilayer graphene.* Nature, 2020. **588**(7839): p. 610-615.
15. Li, G., et al., *Observation of Van Hove singularities in twisted graphene layers.* Nature Physics, 2010. **6**(2): p. 109-113.
16. Andrei, E.Y. and A.H. MacDonald, *Graphene bilayers with a twist.* Nature Materials, 2020. **19**(12): p. 1265-1275.
17. Choi, Y., et al., *Interaction-driven band flattening and correlated phases in twisted bilayer graphene.* Nature Physics, 2021. **17**(12): p. 1375-+.
18. Wong, D.L., et al., *Cascade of electronic transitions in magic-angle twisted bilayer graphene.* Nature, 2020. **582**(7811): p. 198-+.
19. Choi, Y., et al., *Electronic correlations in twisted bilayer graphene near the magic angle.* Nature Physics, 2019. **15**(11): p. 1174-+.
20. Yu, J.C., et al., *Spin skyrmion gaps as signatures of strong-coupling insulators in magic-angle twisted bilayer graphene.* Nature Communications, 2023. **14**(1).
21. Yu, J.C., et al., *Correlated Hofstadter spectrum and flavour phase diagram in magic-angle twisted bilayer graphene.* Nature Physics, 2022. **18**(7): p. 825-+.
22. Zondiner, U., et al., *Cascade of phase transitions and Dirac revivals in magic-angle graphene.* Nature, 2020. **582**(7811): p. 203-+.
23. Xie, Y.L., et al., *Spectroscopic signatures of many-body correlations in magic-angle twisted bilayer graphene.* Nature, 2019. **572**(7767): p. 101-+.
24. Jiang, Y.H., et al., *Charge order and broken rotational symmetry in magic-angle twisted bilayer graphene.* Nature, 2019. **573**(7772): p. 91.
25. Zhou, G.D., et al., *Kondo phase in twisted bilayer graphene.* Physical Review B, 2024. **109**(4).

26. Lau, L.L. and P. Coleman, *Topological mixed valence model for twisted bilayer graphene.* arXiv preprint arXiv:2303.02670, 2023.
27. Hu, H.Y., et al., *Symmetric Kondo Lattice States in Doped Strained Twisted Bilayer Graphene.* Physical Review Letters, 2023. **131**(16).
28. Hu, H.Y., B.A. Bernevig, and A.M. Tsvelik, *Kondo Lattice Model of Magic-Angle Twisted-Bilayer Graphene: Hund's Rule, Local-Moment Fluctuations, and Low-Energy Effective Theory.* Physical Review Letters, 2023. **131**(2).
29. Chou, Y.Z. and S. Das Sarma, *Kondo Lattice Model in Magic-Angle Twisted Bilayer Graphene.* Physical Review Letters, 2023. **131**(2).
30. Kang, J., B.A. Bernevig, and O. Vafek, *Cascades between Light and Heavy Fermions in the Normal State of Magic-Angle Twisted Bilayer Graphene.* Physical Review Letters, 2021. **127**(26).
31. Davenport, J.L., et al., *Probing the electronic structure of graphene near and far from the Fermi level via planar tunneling spectroscopy.* Applied Physics Letters, 2019. **115**(16).
32. Jung, S., et al., *Direct Probing of the Electronic Structures of Single-Layer and Bilayer Graphene with a Hexagonal Boron Nitride Tunneling Barrier.* Nano Letters, 2017. **17**(1): p. 206-213.
33. Chandni, U., et al., *Signatures of Phonon and Defect-Assisted Tunneling in Planar Metal-Hexagonal Boron Nitride-Graphene Junctions.* Nano Letters, 2016. **16**(12): p. 7982-7987.
34. Jung, S., et al., *Vibrational Properties of h-BN and h-BN-Graphene Heterostructures Probed by Inelastic Electron Tunneling Spectroscopy.* Scientific Reports, 2015. **5**.
35. Malec, C.E. and D. Davidovic, *Transport in graphene tunnel junctions.* Journal of Applied Physics, 2011. **109**(6).
36. Dvir, T., et al., *Spectroscopy of bulk and few-layer superconducting NbSe2 with van der Waals tunnel junctions.* Nature Communications, 2018. **9**(1): p. 598.
37. Mohr, M., et al., *Phonon dispersion of graphite by inelastic x-ray scattering.* Physical Review B, 2007. **76**(3).
38. Zhang, Y.B., et al., *Giant phonon-induced conductance in scanning tunnelling spectroscopy of gate-tunable graphene.* Nature Physics, 2008. **4**(8): p. 627-630.
39. Inbar, A., et al., *The quantum twisting microscope.* Nature, 2023. **614**(7949): p. 682-687.
40. Li, G., A. Luican, and E.Y. Andrei, *Scanning Tunneling Spectroscopy of Graphene on Graphite.* Physical Review Letters, 2009. **102**(17): p. 176804.
41. Lu, C.-P., et al., *Local, global, and nonlinear screening in twisted double-layer graphene.* Proceedings of the National Academy of Sciences, 2016. **113**(24): p. 6623-6628.
42. Saito, Y., et al., *Isospin Pomeranchuk effect in twisted bilayer graphene.* Nature, 2021. **592**(7853): p. 220-+.
43. Zhao, Y., et al., *Creating and probing electron whispering-gallery modes in graphene.* Science, 2015. **348**(6235): p. 672-675.
44. Kotliar, G., S. Murthy, and M.J. Rozenberg, *Compressibility Divergence and the Finite Temperature Mott Transition.* Physical Review Letters, 2002. **89**(4): p. 046401.
45. Kwan, Y.H., et al., *Kekul\'e Spiral Order at All Nonzero Integer Fillings in Twisted Bilayer Graphene.* Physical Review X, 2021. **11**(4): p. 041063.

46. Rai, G., et al., *Dynamical Correlations and Order in Magic-Angle Twisted Bilayer Graphene.* Physical Review X, 2024. **14**(3): p. 031045.
47. Saito, Y., et al., *Hofstadter subband ferromagnetism and symmetry-broken Chern insulators in twisted bilayer graphene.* Nature Physics, 2021. **17**(4): p. 478-481.
48. Wagner, G., et al., *Global Phase Diagram of the Normal State of Twisted Bilayer Graphene.* Physical Review Letters, 2022. **128**(15): p. 156401.
49. Lian, B., et al., *Twisted bilayer graphene. IV. Exact insulator ground states and phase diagram.* Physical Review B, 2021. **103**(20): p. 205414.
50. Zou, K., X. Hong, and J. Zhu, *Effective mass of electrons and holes in bilayer graphene: Electron-hole asymmetry and electron-electron interaction.* Physical Review B, 2011. **84**(8).
51. Rafael Luque Merino, D.C., Haoyu Hu, Jaime Diez-Merida, Andres Diez-Carlon, Takashi Taniguchi, Kenji Watanabe, Paul Seifert, B. Andrei Bernevig, Dmitri K. Efetov, *Evidence of heavy fermion physics in the thermoelectric transport of magic angle twisted bilayer graphene.* arXiv:2402.11749 2024.
52. Bultinck, N., et al., *Ground State and Hidden Symmetry of Magic-Angle Graphene at Even Integer Filling.* Physical Review X, 2020. **10**(3): p. 031034.
53. Bernevig, B.A., et al., *Twisted bilayer graphene. V. Exact analytic many-body excitations in Coulomb Hamiltonians: Charge gap, Goldstone modes, and absence of Cooper pairing.* Physical Review B, 2021. **103**(20): p. 205415.
54. Qianying Hu, S.L., Xinheng Li, Hao Sh, Xi Dai and Yang Xu, *Link between cascade transitions and correlated Chern insulators in magic-angle twisted bilayer graphene.* arXiV: 2406.08734, 2024.
55. Călugăru, D., et al., *Twisted bilayer graphene as topological heavy fermion: II. Analytical approximations of the model parameters.* Low Temperature Physics, 2023. **49**(6): p. 640-654.
56. Dumitru Călugăru, H.H., Rafael Luque Merino, Nicolas Regnault, Dmitri K. Efetov and B. Andrei Bernevig, *The Thermoelectric Effect and Its Natural Heavy Fermion Explanation in Twisted Bilayer and Trilayer Graphene.* arXiv: 2402.14057, 2024.
57. Xie, Y., et al., *Spectroscopic signatures of many-body correlations in magic-angle twisted bilayer graphene.* Nature, 2019. **572**(7767): p. 101-105.
58. Nuckolls, K.P., et al., *Quantum textures of the many-body wavefunctions in magic-angle graphene.* Nature, 2023. **620**(7974): p. 525-532.
59. Jonah Herzog-Arbeitman. Dumitru Călugăru, H.H., Jiabin Yu, Nicolas Regnault, Jian Kang, B. Andrei Bernevig and Oskar Vafek, *Kekul\'e Spiral Order from Strained Topological Heavy Fermions.* arXiv:2502.08700, 2025.
60. Hu, H., B.A. Bernevig, and A.M. Tsvelik, *Kondo Lattice Model of Magic-Angle Twisted-Bilayer Graphene: Hund's Rule, Local-Moment Fluctuations, and Low-Energy Effective Theory.* Physical Review Letters, 2023. **131**(2): p. 026502.
61. Herzog-Arbeitman, J., et al., *Heavy Fermions as an Efficient Representation of Atomistic Strain and Relaxation in Twisted Bilayer Graphene.* arXiv preprint arXiv:2405.13880, 2024.
62. Ledwith, P.J., et al., *Nonlocal Moments in the Chern Bands of Twisted Bilayer Graphene.* arXiv e-prints, 2024: p. arXiv:2408.16761.

63. J. A. Leon, N.C.M., A. Rahim, L. E. Gomez, M. A. P. da Silva, G. M. Gusev, *Transferring Few-Layer Graphene Sheets on Hexagonal Boron Nitride Substrates for Fabrication of Graphene Devices.* Graphene, 2014. **3**.
64. Kim, K., et al., *van der Waals Heterostructures with High Accuracy Rotational Alignment.* Nano Letters, 2016. **16**(3): p. 1989-1995.
65. Britnell, L., et al., *Electron Tunneling through Ultrathin Boron Nitride Crystalline Barriers.* Nano Letters, 2012. **12**(3): p. 1707-1710.
66. Wang, L., et al., *One-Dimensional Electrical Contact to a Two-Dimensional Material.* Science, 2013. **342**(6158): p. 614-617.
67. Britnell, L., et al., *Field-Effect Tunneling Transistor Based on Vertical Graphene Heterostructures.* Science, 2012. **335**(6071): p. 947-950.
68. Shiga, K., et al., *Electrical transport properties of gate tunable graphene lateral tunnel diodes.* Japanese Journal of Applied Physics, 2020. **59**(SI): p. SIID03.
69. Yang, H., et al., *Graphene Barristor, a Triode Device with a Gate-Controlled Schottky Barrier.* Science, 2012. **336**(6085): p. 1140-1143.

**Extended Data**

**Methods**

**Sample fabrication**

The planar tunneling stacks were prepared using the dry transfer method[63] and the 'tear and stack' method[64] in a dry argon atmosphere inside a glovebox. A narrow graphite ribbon (less than 5 um wide) was first exfoliated onto a PMMA film and then dry-transferred onto a bottom h-BN flake (30-70 nm thick) exfoliated on a $SiO_2$/Si substrate. An ultra-thin h-BN sheet (2-6 layers thick) serving as the tunneling layer[65] was then exfoliated and dry-transferred to cover one end of the graphite probe. The remaining area was covered with another cushion h-BN flake (5-10 nm thick) to prevent unwanted tunneling, forming a probing area of ~1 μm × 1 μm. The entire bottom stack was annealed at 300 °C in forming gas (10% $H_2$ and 90% Ar) overnight to remove PMMA residue. The top stack was constructed by first picking up a top h-BN flake (30-50 nm thick) using a PDMS/PPC hemispherical handle. A large monolayer graphene, exfoliated on $SiO_2$/Si substrate, was then picked up in halves by the h-BN flake with a twist angle of 1.1° between two layers. The h-BN/TBG top stack was finally released onto the bottom stack at around

90°C, aligning the bubble-free TBG region with the small tunneling probe. An Au top gate was deposited over the TBG area for electrical gating. Electrical edge contacts (Cr/Au) were deposited after e-beam lithography and etching with $CHF_3$ plasma[66]. The temperature during fabrication was kept below 160°C to prevent possible relaxation of TBG.

**Planar tunneling measurements**

The planar tunneling measurements were performed in a $^3$He refrigerator at a base temperature of 0.3 K, with a maximum perpendicular magnetic field of 8 T. The TBG was electrically gated using a gate voltage $V_g$ applied to the Au top gate relative to the sample, and the tunneling current was driven by a bias voltage $V_b$ applied to the sample relative to the tunneling probe. Differential conductance curves were measured using the standard lock-in technique, with an AC bias voltage modulation (0.2-1 mV, 11.07 Hz) superimposed on the DC $V_b$, or with an AC gate voltage modulation (5-10 mV, 11.07 Hz) superimposed on the DC $V_g$. The tunneling current was amplified by an external current preamplifier with a gain of $10^9$ V/A and measured at the same frequency as the AC voltage modulation.

**Carrier density per moiré cell and twist angle**

To accurately determine the carrier density and twist angle we use high magnetic field Landau levels (LLs) and Chern insulators (CIs) (Fig, 2c). Using the LL fan at the CNP, $n(V_g) = \nu(B/\phi_0)$ where $\nu = 4, 8, 12 ...$ is the LL filling factor and $\phi_0$ the flux quantum, we obtain $n = 0.30 \times 10^{12} V_g$ (cm$^{-2}$). From this relation, together with the trajectories of Chern insulators which follow the Streda formula $n(V_g) = sn_0 + \nu(B/\phi_0)$ where $s, \nu \in \mathbb{Z}$ and, we obtain $n_0 = 0.67 \times 10^{12}$ cm$^{-2}$ and the twist angle in radians $\theta = a(\frac{2}{\sqrt{3}n_0})^{\frac{1}{2}}$, which corresponds to 1.07°.

**Normalized tunneling current and differential conductance**

We analyze the measured tunneling current, $I = I_0 \int_{E_F}^{E_F+eV_b} T(n)\rho(\epsilon)d\epsilon$, under the assumption that the probe DOS and transmission coefficient, $T(n)$, are energy-independent [27, 31, 34, 37]. We further simplify the expression by dividing it by an exponential transmission coefficient background $T(n) \sim \exp(-\frac{2d}{\hbar}\sqrt{2m\Delta(n)})$ [67-69] (SI) resulting from the barrier height dependence on carrier density, $\Delta(n)$. This gives the normalized tunneling current $I_n = I/T(n) = I_0 \int_{E_F}^{E_F+eV_b} \rho(\epsilon)d\epsilon$, where $I_0$ is a constant. Taking the derivative with respect to $V_b$, and recalling that $E_F = \mu(n)$, we obtain

$$\frac{dI_n}{dV_b} = I_0[(\frac{\partial(\mu(n)+eV_b)}{\partial V_b})\rho(\mu(n)+eV_b) - \frac{\partial \mu(n)}{\partial V_b}\rho(\mu(n))]$$

$$= eI_0\left[\rho(E_F+eV_b) - \frac{d\mu(n)}{dn}\frac{C_b}{e^2}(\rho(E_F+eV_b) - \rho(E_F))\right] \quad (1)$$

Where $C_b = \frac{e\partial n}{\partial V_b} \sim 1.3$ uF/cm$^2$ is the capacitance between the sample and the bias electrode.

In our planar tunneling structure, the ultra-stable tunneling junction enables us to measure the derivative of the tunneling current with respect to the gate voltage ($V_g$)

$$\frac{dI_n}{dV_g} = I_0[\frac{\partial \mu(n)}{\partial V_g}\rho_s(\mu(n)+eV_b) - \frac{\partial \mu(n)}{\partial V_g}\rho_s(\mu(n))]$$

which simplifies to:

$$\frac{dI_n}{dV_g} = eI_0\frac{d\mu(n)}{dn}\frac{C_g}{e^2}\left(\rho_s(\mu(n)+eV_b) - \rho_s(\mu(n))\right). \quad (5)$$

By solving equations (4) and (5), we arrive at the following expressions:

$$\rho_s(\mu(n)+eV_b) = (\frac{dI_n}{dV_b} + \frac{C_b}{C_g}\frac{dI_n}{dV_g})/eI_0, \quad (6)$$

$$\frac{d\mu(n)}{dn} = \frac{e\frac{dI_n}{dV_g}\frac{1}{I_0 C_g}}{\rho_s(\mu(n)+eV_b)-\rho_s(\mu(n))}, \tag{7}$$

where $C_g$ is the capacitance between the sample and the gate electrode.

**Determination of effective mass at CNP and remote bands through Landau levels**

Due to suppressed kinetic energy and small bandwidth of the flat bands, Landau levels (LLs) in these bands are closely spaced and difficult to observe in energy spectra. However, from our planar tunneling results --- specifically, the high d$\mu$/dn and the appearance of dark lines under perpendicular magnetic fields --- we identify elevated d$\mu$/dn at doping levels corresponding to gaps between LLs, with a Landau filling factor $\nu = 0, \pm 4, \pm 8, \pm 12$, and a degeneracy of 4 near the CNP. This observation is consistent with transport measurements that show $R_{xx}$ dips and quantized $R_{xy}$. In contrast, the remote bands, which have a larger Fermi velocity, are expected to display more pronounced LLs. This is evident in the d$I$/d$V_b$ map under a magnetic field of 6T (Fig.S4 in SI).

To study the LLs of the flat bands in detail, we conducted high-resolution d$I$/d$V_b$($V_b$, $n/n_0$) map measurements at a base temperature of 0.3 K near the CNP. The Landau filling factor $\nu$ can be calculated using the Streda formula: $n/n_0(s,\nu) = s + \nu\,(\phi/\phi_0)$ where $s, \nu \in \mathbb{Z}$. In Extended Fig.2a, 2b where the d$I$/d$V_b$ is plotted at a fixed $V_b$ from 0 T to 2 T, LLs with $\nu = \pm 12, \pm 8, \pm 4, 0$ emerge at fields as low as 0.6 T. Above 1.5 T, all the degeneracies of the 0$^{\text{th}}$ LL are lifted, consistent with the previous transport measurements[13].

The differential conductance with background subtraction d$I$/d$V_b^*$, is plotted as a function of $V_b$-$V_D$ ($V_D$ is the bias voltage at the Dirac point) from 1 T to 3 T, highlighting LLs on the electron side (Extended Fig. 2c). The background is defined as an average of the $dI/dV_b$ spectra over the Landau filling $\nu$. The energy difference between LLs follows a linear relation with the magnetic field, characteristic of massive Dirac fermions, rather than the square root dependence typical of massless Dirac fermions. Both the Landau filling sequence and this linear dependence support the presence of massive chiral bands near the CNP. By fitting the LLs of the flat bands to the energy expression for massive Dirac fermion $E_N - E_0 = \hbar\omega_c\sqrt{N(N-1)}$ we extract an effective mass $m^* = 0.174\ m_e$.

A similar analysis of the LLs for remote bands is shown in Extended Fig. 2d. Near the CNP, LLs for those bands appear at higher energies, starting at 60 mV above 1 T. The bias voltage of the LL peaks for both flat and remote bands as a function of $\sqrt{N(N+1)}B$ is shown in Extended Fig.2e. From the energy expression for the $N^{th}$ LL relative to the Dirac point $E_N - E_0 = \hbar\omega_c\sqrt{N(N-1)}$, where $\omega_c = eB/m^*$, the slope is inversely proportional to the effective mass. The remote bands have a much lighter $m^* = 0.047\ m_e$ than the flat bands, consistent with the expectation of dispersive remote bands and flat central bands. However, note that this estimation of $m^*$ for flat bands applies to low fillings near the CNP. As the system crosses integer fillings, it undergoes strong transitions from heavy to light effective mass.

**Extraction of entropy from T-dependent chemical potential**

The entropy is calculated from temperature-dependent chemical potential results (Extended Fig.3a) by integrating the Maxwell's relation

$$\left(\frac{\partial S}{\partial \nu}\right)_T = -\left(\frac{\partial \mu}{\partial T}\right)_\nu, \tag{11}$$

where $S$ is the entropy per moiré unit cell, $\mu$ is the chemical potential, $T$ is the temperature, and $\nu = n/n_0$ is the moiré filling factor. Similar to previous single-electron transistor experiments[11, 42], we fix the chemical potential at the CNP to 0 and assume that the entropy at filling of 4 is 0 due to the large band gap separating the flat bands from the remote bands.

$$S\left(\frac{n}{n_0}, T\right) = -\int_4^{\frac{n}{n_0}} \left(\frac{\partial \mu}{\partial T}\right)_\nu d\nu. \tag{12}$$

To minimize the error, we switch the order of integration and differentiation, which does not affect the entropy calculation:

$$S\left(\frac{n}{n_0}, T\right) = -\frac{\partial \int_4^{\frac{n}{n_0}} \mu_\nu d\nu}{\partial T}. \quad (13)$$

The integral of the chemical potential $-\int_4^{n/n_0} \mu_\nu d\nu$ for fillings of 2 and 3, from 4 K to 40 K is shown in Extended Figure 4b. The entropy values and corresponding errors are obtained by performing a linear fit using three adjacent temperature points, as displayed in Extended Figure 4c.

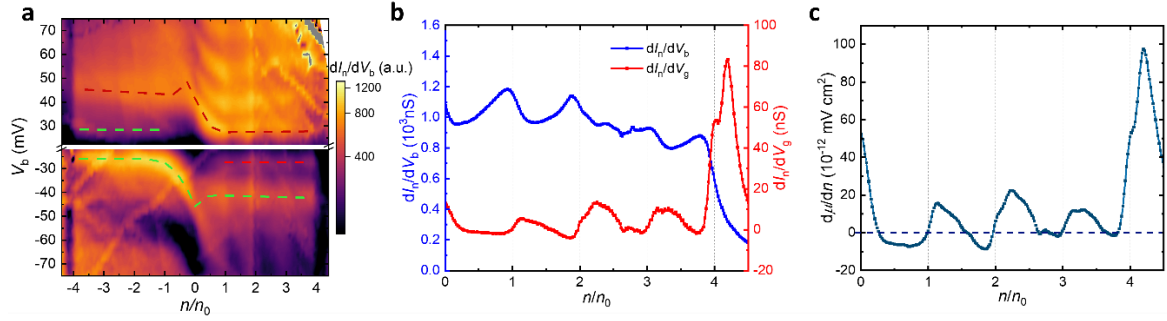

**Extended Fig. 1 dI$_n$/dV$_b$ map and Extraction of inverse compressibility. a,** dI$_n$/dV$_b$ map as a function of $V_b$ and $n/n_0$. The phonon-induced gap feature near $V_b = 0$ mV is omitted. **b,** dI$_n$/dV$_b$ and dI$_n$/dV$_g$ on the electron side taken at $V_b = -42$ mV at 4 K. **c,** Extracted d$\mu$/d$n$ on the electron side at 4 K with data in **a**.

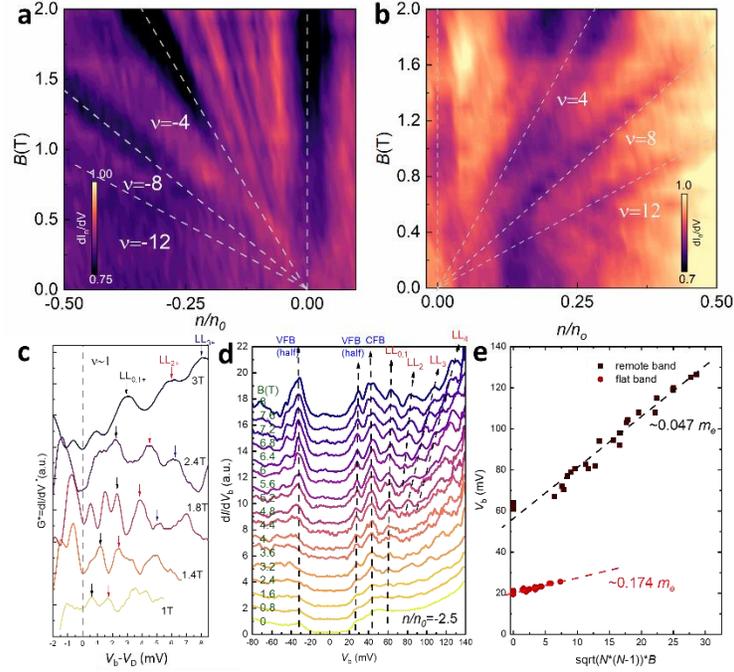

**Extended Fig.2 Landau levels of flat bands and remote bands. a**, High-resolution d$I_n$/d$V_b$(40 mV) near CNP on the hole side. The low d$I_n$/d$V_b$ dark lines correspond to the high compressibility of gaps between LLs. **b**, Same as **a**, but measured at -40 mV and on the electron side. **c**, d$I$/d$V_b$*(stacked) as a function of $V_b$-$V_D$ ($V_D$ is the bias voltage of the Dirac point) at the fixed $\nu\sim1$ near the CNP from 1 to 3 T. The LLs of the flat bands are labeled with arrows. **d**, d$I$/d$V_b$ (stacked) as a function of $V_b$ at $n/n_0$ = -2.5 from 0 to 8 T. The flat bands (VFB: valence flat band, CFB: conduction flat band) and the LLs of remote bands are labeled with dashed lines. **e**, $V_b$ of the LL peaks as a function of $\sqrt{N(N-1)B}$ taken from **c** and **d**. $E_N - E_0 = \hbar\omega_c\sqrt{N(N-1)}$ for massive chiral bands, $\omega_c = eB/m^*$. The linear fittings claim an effective mass of 0.047 $m_e$ and 0.174 $m_e$ for remote bands and flat bands, respectively.

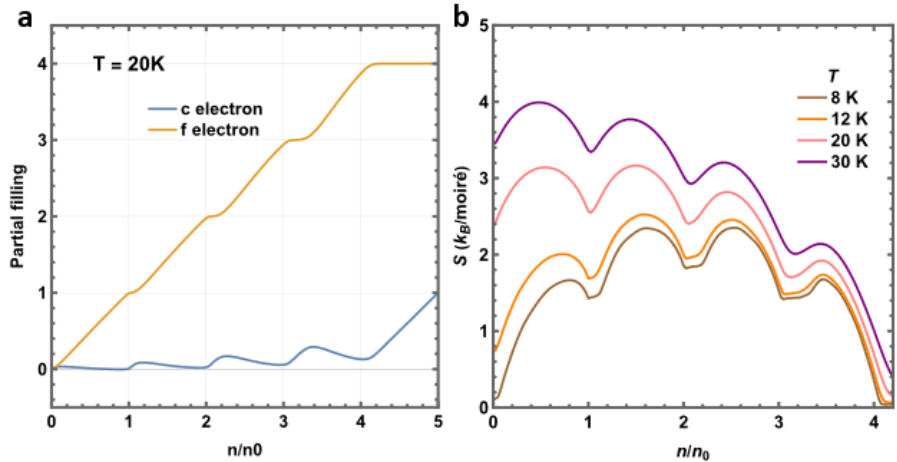

**Extended Figure 3 Partial filling and entropy $S$. a**, Partial filling of $f$- and $c$-electrons calculated in the atomic limit of the topological heavy-fermion model with strain. The Hubbard interaction energy $U_1$ of 23meV and the compressive heterostrain $\varepsilon=0.001$ are set to match the experimental data (see SI). Within the flat bands, most electrons are localized $f$-electrons at AA sites. Above full filling, mobile c electrons have high occupancy in the remote bands. **b**, Total entropy $S$ on the electron side at various temperatures calculated with the same parameters as **a**. At low temperatures below 10 K, the non-interacting conduction and valence flat bands separate due to heterostrain, each with degeneracy of 4. As temperature goes up, charge transfer becomes possible between flat bands and the degeneracy gets doubled to 8, increasing the total entropy.

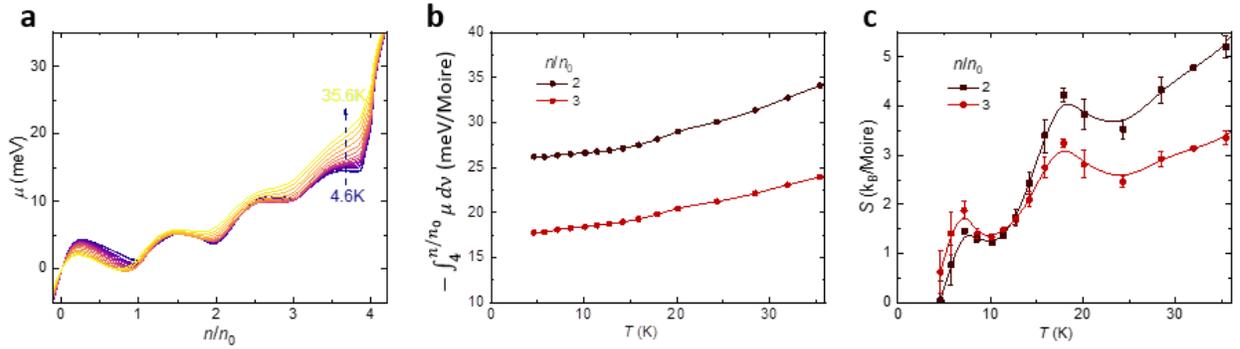

**Extended Figure 4 Extraction of entropy. a**, Chemical potential on the electron side from 4 K to 40 K. The chemical potential at CNP is set to 0. **b**, The integrals of chemical potential at filling of 2 and 3 as a function of temperature. **c**, Extracted entropy with the formula $S\left(\frac{n}{n_0}, T\right) = -\frac{\partial \int_4^{n/n_0} \mu_\nu d\nu}{\partial T}$ and data in **b**. The entropy curves show two plateaus around 10 K and 20 K.